\newcommand{\PRE}[1]{}       
\documentclass[aps,prl,twocolumn,preprintnumbers,
              showpacs,nofootinbib]{revtex4}

\usepackage{bm}
\usepackage{graphicx}

\graphicspath{{figs/}}

\newcommand{\gev}{\text{GeV}}

\newcommand{\m}{\text{m}}

\newcommand{\eqref}[1]{Eq.~(\ref{#1})}

\newcommand{\be}{\begin{equation}}
\newcommand{\ee}{\end{equation}}
\newcommand{\bea}{\begin{eqnarray}}
\newcommand{\eea}{\end{eqnarray}}
\newcommand{\gsim}{\lower.7ex\hbox{$\;\stackrel{\textstyle>}{\sim}\;$}}
\newcommand{\lsim}{\lower.7ex\hbox{$\;\stackrel{\textstyle<}{\sim}\;$}}

\newcommand{\ssection}[1]{{\em #1.\ }}

\begin{document}

\preprint{MIFPA-10-52}
\preprint{UH-511-1156-2010}

\title{
\PRE{\vspace*{1.3in}}
Asymmetric Dark Matter from Hidden Sector Baryogenesis
\PRE{\vspace*{0.3in}}
}

\author{Bhaskar Dutta}
\affiliation{Department of Physics and Astronomy,
Mitchell Institute for Fundamental Physics,
Texas A\&M University, College Station, TX  77843, USA
}

\author{Jason Kumar}
\affiliation{Department of Physics and Astronomy, University of
Hawai'i, Honolulu, HI 96822, USA
}


\begin{abstract}
\PRE{\vspace*{.3in}} We consider the production of asymmetric dark matter
during hidden sector baryogenesis.  We consider a particular supersymmetric model where
the dark matter candidate has a number density approximately equal to the
baryon number density, with a mass of the same scale as the $b$, $c$ and $\tau$.
Both baryon asymmetry and dark matter are created at the same time in this model.
We describe collider and direct detection signatures of this model.
\end{abstract}

\maketitle

\ssection{Introduction}
There are two remarkable coincidences which have motivated many
theoretical models of dark matter.  The first is the fact that the
dark matter density is approximately the same (within an order of magnitude)
as what one would expect from a stable thermal relic at the weak scale,
and the second is that fact that the dark matter density is approximately
the same the baryon density.  Thermal WIMPs and WIMPless dark matter~\cite{Feng:2008ya}
are examples of models which utilize the first coincidence to explain the
observed dark matter density.  Models which utilize the second one could contain
asymmetric dark matter~\cite{asymDM} or non-thermal dark matter~\cite{Allahverdi:2010rh}.

Models of asymmetric dark matter rely on the fact that a stable particle
with the same mass and number density as baryonic matter would have
roughly the mass density to explain our cosmological observations of
dark matter.  To utilize this coincidence, one must explain why the dark
matter particle has a mass $m_{DM} \sim {\cal O} (\gev )$, and why the number density
is similar to that of baryons.  Many such models thus tie the mechanism of
generating dark matter to baryogenesis.

We will consider the possibility of generating a dark matter candidate
utilizing hidden sector baryogenesis~\cite{Dutta:2006pt}.  We will find that we naturally
get a dark matter candidate with about the same number density as baryons.
The model we consider is in the framework of supersymmetry, and 
both the baryon asymmetry and dark matter are created at the same time.
Furthermore, we will find that we can easily accommodate $m_{DM} \sim {\cal O}(\gev )$,
and that this choice is correlated with the mass scale of the bottom and
charm quarks, and the tau lepton. We also discuss signals at the
Large Hadron Collider (LHC).

The organization of this paper is as follows. We first review hidden sector baryogenesis.
We then discuss the mass scale of the new particles and the
asymmetric dark matter candidate. After that,
we discuss possible flavor constraints, and direct detection and collider signals. We
close the paper with concluding remarks.

\ssection{Review of hidden sector baryogenesis}
Hidden sector baryogenesis is a generalization of the idea behind
electroweak baryogenesis~\cite{Dutta:2006pt}.
The idea of hidden sector baryogenesis is that sphalerons of a hidden
sector gauge group can generate a baryon asymmetry in the Standard
Model sector.  
We will formulate this as a supersymmetric model.
The setup we will seek is a hidden sector gauge group
$G$, with chiral matter charged under both $G$ and $SU(3)_{QCD} \subset U(3)$
(the diagonal $U(1)$ subgroup of this $U(3)$ will be $U(1)_B$, whose charge
is baryon number).
We assume that $G$ has a diagonal subgroup $U(1)_G$.
We will denote by $q_i$ an exotic quark multiplet which is charged under
the fundamental of $G$ and under $U(1)_B$ (and thus also $SU(3)_{QCD}$), but not $SU(2)_L$.
There is thus a $U(1)_B G^2$ mixed anomaly,
implying that the divergence of the baryon current is
\bea
\partial_{\mu} j_B^{\mu} &\propto& {1\over 32\pi^2} (g_G ^2 Tr\,F_G \wedge F_G  +... ).
\eea
We then see that sphaleron or instanton effects in the hidden $G$ sector can
generate a configuration such that the right side of the above equation is
non-zero.  This implies a non-zero divergence of the baryon current, resulting
in a change in baryon number.
If the $G$ gauge group breaks through a strongly first-order
phase transition, and if there is sufficient $CP$-violation at the domain
wall of the phase transition, then a baryon asymmetry can be generated
at the phase transition.  This asymmetry takes the form of a flux of $q_i$ exotic
quarks from the domain wall.  Note that this asymmetry can be generated regardless
of the mass scale at which $G$ breaks, provided the universe is at some point
hot enough to be in the phase of unbroken $G$-symmetry.

Eventually, these $q_i$ quarks must decay to Standard Model quarks.
Thus, it is not sufficient to add only the exotic quarks $q_i$.   An additional multiplet
can be added to permit the decay $q_i \rightarrow q_{SM(R)} \tilde \eta$, where $q_{SM(R)}$ is a
right-handed Standard Model quark and $\tilde \eta$ is the scalar component of a
supermultiplet\footnote{We follow a convention where the same letter is used to denote a supermultiplet
and its fermionic component, while a tilde denote the scalar component.}
which is also
charged under $G$, but is neutral under $SU(3)_{qcd}$ and $U(1)_Y$.
$U(1)_{T3R}$ is a group under which some right-handed fermions
are charged, analogous to the $U(1)_{T3L}$ subgroup of the electroweak $SU(2)$
under which left-handed Standard Model fermions are charged.

In addition to $G$ and $U(1)_{T3R}$, this model contains an
additional symmetry (taken for simplicity to be a discrete $Z_2$); the lightest particle
charged under this symmetry is thus stable.
The relevant matter content of this model is given in Table I
for the case where the new exotic quark is up-type.  If it is down-type,
the matter content of an example model is given in Table II.

\begin{table}
\begin{center}
\caption{Particle spectrum for an example model with an
exotic up-type quark.
$Q_G$ is the charge under $U(1)_G$, the diagonal subgroup of
$G$.
The matter content below the line
is already present in the Standard Model.}
\begin{tabular}{|r|r|r|r|r|r|}
  \hline
  supermultiplet & $Q_B$ & $Q_G$ & $Q_{T_{3R}}$ & $Q_Y$ & $Z_2$ \\
  \hline
  $q_i$ & ${1\over 3}$ & -1 & 0 & ${2\over 3}$ & $-$\\
  $q'_i$ & $-{1\over 3}$& 0 & 0 & $-{2\over 3}$ & $-$\\
  $\eta$ & 0 & 1 & 1  &0 & $-$\\
  $\eta'$ & 0 & -1 & 0  &0 & $-$\\
  $\xi_j$ & 0 & -1 & 0 & $-{2\over 3}$ & $+$\\
  $\xi'_j$ & 0 & 0 & 0 & ${2\over 3}$ & $+$ \\
  \hline
  $b_R$ & $-{1\over 3}$& 0 & 1 &  ${1\over 3}$ & $+$\\
  $c_R$ & $-{1\over 3}$& 0 & -1 &  $-{2\over 3}$ &$+$\\
  $\tau_R$ & 0& 0 & 1 &  1 & $+$\\
  \hline
\end{tabular}
\end{center}
\label{mattercontentup}
\end{table}
\begin{table}
\label{mattercontentdown}
\begin{center}
\caption{Similar to Table I, the particle
spectrum for an example model with an
exotic down-type quark.}
\begin{tabular}{|r|r|r|r|r|r|}
  \hline
  supermultiplet & $Q_B$ & $Q_G$ & $Q_{T_{3R}}$ & $Q_Y$ & $Z_2$ \\
  \hline
  $q_i$ & ${1\over 3}$ & 1 & 0 & $- {1\over 3}$ & $-$\\
  $q'_i$ & $-{1\over 3}$& 0 &  0 & ${1\over 3}$ & $-$ \\
  $\eta$ & 0 & -1 & -1 &0 & $-$\\
  $\eta'$ & 0 & 1 & 0 &0 & $-$\\
  $\xi_j$ & 0 & -1 & 0 &  ${1\over 3}$ & $+$ \\
  $\xi'_j$ & 0 & 0 & 0 &  $-{1\over 3}$ & $+$ \\
  \hline
  $b_R$ & $-{1\over 3}$& 0 & 1 & ${1\over 3}$ & $+$\\
  $c_R$ & $-{1\over 3}$& 0 & -1 & $-{2\over 3}$ & $+$\\
  $\tau_R$ & 0& 0 & 1  & 1 & +\\
  \hline
\end{tabular}
\end{center}
\end{table}

It is necessary for all cubic anomalies and the hypercharge mixed anomaly to cancel.
This cancelation must be manifest for the matter content with mass below the electroweak
symmetry breaking scale.  The anomalies induced by heavier matter must also
cancel to keep the photon massless, and for this purpose the $\xi_j$, $\xi'_j$ multiplets are also
added (with 3 times the multiplicity of the $q_i$).  There is little experimental constraint
on the hidden sector matter which can exist well above the electroweak symmetry-breaking
scale, so we have considerable freedom in satisfying these anomaly constraints.    We need only demand
that, although hypercharge anomalies cancel, there is a $U(1)_B G^2$ mixed anomaly.
Sphalerons of the group $G$ can thus generate non-vanishing baryon charge, but not
$G$-charge or hypercharge.

Since $\tilde \eta$ is electrically neutral and a singlet under $SU(3)_{qcd}$, it is a potential
dark matter candidate (as is the fermionic partner of the multiplet, which we will denote as $\eta$).
Moreover, as we will see, the number density of the lightest component of the $\eta$ multiplet
is proportional to the baryon number density, and its mass is $\sim {\cal O}(\gev )$.

\ssection{Mass-scale of the particle content}
The mass scales of the new particles of this model are determined by the
symmetry-breaking scales of the two new continuous gauge groups, $G$ and
$U(1)_{T3R}$.
For the $U(1)_Y U(1)_{T3R}^2$ mixed anomalies to vanish for light matter content,
there must be an up-type quark, a down-type quark and a charged lepton which are all
charged under $U(1)_{T3R}$.  But these fermions do not in fact have to be in the same
generation.
Since the right-handed quarks
are charged under $U(1)_{T3R}$ but the left-handed quarks are not,
one would expect that the mass scale of those
fermions is set by the symmetry-breaking scale of $U(1)_{T3R}$.
Since the $b$, $c$ and $\tau$ all have masses of approximately the same scale,
we will choose the right-handed $c$ and $b$ quarks and the right-handed $\tau$ as
the fermions charged under $U(1)_{T3R}$.  The mass scale of these Standard Model
fermions naturally suggests that $U(1)_{T3R}$
should have a symmetry-breaking scale of ${\cal O}(\gev )$.

We can model $U(1)_{T3R}$ symmetry breaking through
a pair of ``higgs-like" scalars, which we will denote as $\tilde \phi_{u,d}$
(we will denote their fermionic partners as $\phi_{u,d}$), with charges
$\pm 1$ under $U(1)_{T3R}$ which get vacuum expectation values.
Similarly, we can model symmetry breaking of $G$ by a higgs-like scalar $\tilde \Phi_G$,
which is charged under $G$ and whose vacuum expectation value spontaneously breaks $G$.
At low energies (well below the
scale of electroweak symmetry breaking), the mass terms of the bottom and charm
quarks and the tau lepton are then controlled by the vevs of $\tilde \phi_{u,d}$.  We
may thus write the following Yukawa couplings:
\bea
\label{RHfermmass}
V_{mass} &=& \lambda_b \tilde \phi_d \bar b_L b_R +\lambda_c \tilde \phi_u \bar c_L c_R
+\lambda_\tau \tilde \phi_d \bar \tau_L \tau_R +h.~c.
\nonumber\\
m_b &=& \lambda_b \langle \tilde \phi_d \rangle
\nonumber\\
m_c &=& \lambda_c \langle \tilde \phi_u \rangle
\nonumber\\
m_\tau &=& \lambda_\tau \langle \tilde \phi_d \rangle
\eea
If $\langle \tilde \phi_{u,d} \rangle \sim {\cal O}(\gev )$, then we would need
$\lambda_{b,c,\tau} \sim {\cal O}(1)$ in order to get the measured masses of
$b$, $c$ and $\tau$.

One expects the natural mass scale of any new particle to be
set approximately by the lightest symmetry-breaking scale of the groups
under which the particle is chirally charged.  Since the exotic quarks
are not charged under $SU(2)_L$, their mass is not controlled by
electroweak symmetry-breaking.
Instead, $q_i$ and $q'_i$ can obtain mass through a potential term of the form
$\lambda_q \langle \tilde \Phi_G \rangle \bar q'_i q_i$,
we find $m_{q_i} \sim \lambda_q \langle \tilde \Phi_G \rangle$,
and the natural mass scale of the exotic quarks is the symmetry breaking scale of $G$.
There is no a priori constraint on this symmetry breaking scale, but the non-observation
of exotic quarks implies that $G$ breaks at a scale larger than a few hundred GeV.

Similarly, since the $\eta$ and $\eta'$ supermultiplets are vectorlike under $G$ but
chiral under $U(1)_{T3R}$,  they can obtain mass through a
superpotential term of the form $\lambda_\eta \langle \tilde \phi_{u,d} \rangle \bar \eta' \eta$.
The mass of particles in these supermultiplets is thus set by the symmetry-breaking scale
of $U(1)_{T3R}$, which is $\sim {\cal O}(\gev )$.

\ssection{Dark matter asymmetry}
Sphalerons/instantons of the $G$ group generate $q_i$, $\eta$ and also the
$G$-charged matter $\xi_j$; the number densities of the generated particles are
thus correlated.
Since $\xi_i$ is neutral under $Z_2$, its decays will not produce a dark matter asymmetry.
The $q_i$ are charged under $Z_2$, so its decays are mediated by the Yukawa coupling
\bea
W_{yuk.} &=& C_{i} q_i (b,c)_R  \eta  +...
\eea
(depending on whether the exotic quark is down or up type).
The decay $q_i \rightarrow (b,c)_R \tilde \eta$ is kinematically
allowed, since we expect the mass of $q_i$ to be relatively high (set by the symmetry-breaking
scale of $G$), while $m_\eta \sim {\cal O}(\gev )$.

All of the particles generated by sphalerons/instantons of $G$ thus decay to either
$\eta$ or $\tilde \eta$, and the heavier of these will eventually decay to the lighter one.
Since the $\tilde \eta$ and $\tilde \eta^*$ can annihilate efficiently (e.g., via $Z_R$ in $s$-channel),
we are left with the asymmetric component of dark matter density, i.e.,
${\rm \# density_{\eta^*}} \ll {\rm \# density_{\eta}}$.
We thus find that this model gives us exactly what we were looking for, a dark matter
candidate $\tilde \eta$ ($\eta$) with a mass $\sim {\cal O} (\gev )$ and with a number density proportional to the
baryon number density.  The lightest particle of the $\eta$ supermultiplet
is therefore a good asymmetric dark matter candidate.

$G$-spahelerons/instantons would produce an $\bar \eta$ number density which is about
${1\over 3}$ the $q_i$ number density (which we can see from the matter
content).  The decay of
the $q_i$ to Standard Model quarks produce 3 $\tilde \eta$ for each Standard Model hadron.
Assuming that the $\bar \eta$ decay to $\tilde \eta^*$, and that
electroweak sphalerons will convert approximately half of the baryon number density into
leptons, we get a number density ratio ${\rm \# density_{\tilde \eta} \over \# density_{proton}}\sim  4$.
If $\m_{\tilde \eta} \sim {\cal O}(\gev )$,  then we would
have about the right relic density. Both baryon asymmetry and dark matter are created at the same time.

If dark matter annihilation has not frozen out by the time the baryon asymmetry is generated, then
dark matter self-annihilation can wash out any dark matter asymmetry generated by hidden sector baryogenesis
at the $G$ symmetry-breaking phase transition.  Since the symmetry-breaking scale of $G$ is
greater than a few hundred GeV (and thus much greater than the dark matter mass), is likely that dark matter
self-annihilation would not have frozen out by the time of hidden sector baryogenesis, and must be suppressed
in some other way.
But dark matter self-annihilation can be easily forbidden if the $\eta$ supermultiplet is charged under
an unbroken continuous symmetry (either global or gauged).  We will thus assume that the dark matter is
charged under some other such continuous global symmetry; this symmetry will play no role in the
remainder of the discussion.

\ssection{Flavor constraints}
Because the new matter only couples to one generation, it does not  induce flavor changing neutral currents
through renormalizable operators.  FCNC's
can be introduced through non-renormalizable operators of the form $\lambda_{ij} h \tilde \phi \bar f_{Li}
f_j$; where $h$  is the SM Higgs. These may provide an interesting signature for these models, but no current constraint
(since the coefficients may be small).
The main experimental constraint on this model then comes from the process
$b{\bar b} \rightarrow \tilde \phi_d , Z_R \rightarrow \tau \bar \tau$
($Z_R$ is the gauge-boson of
$U(1)_{T3R}$).  This process violates lepton universality, and is
bounded by data from B-factories, such as Belle and BaBar,
at the 0.1\% level~\cite{Lusiani:2007cb} using
searches for $\Upsilon$ decay to $\tau \bar \tau$ pairs.
If the coupling constant of $U(1)_{T3R}$ is small, the exchange
of the $Z_R$ gauge boson may be negligible.  But the exchange of $\tilde \phi_d$
cannot be arbitrarily small, since the couplings $\lambda_{b,\tau}$ are expected to be
of ${\cal O}(1)$ in order to naturally explain the mass scale of the $b$ and $\tau$.

Assuming no accidental coincidence between the mass of an $\Upsilon$ resonance and the
mediating particle, the amplitude for $b \bar b \rightarrow \tau \bar \tau$ is inversely proportional to the
squared mass of the mediating particle (or of the dark matter, when mediated by
a photon).  But since the masses of $Z_R$, $\tilde \phi_d$ and the dark matter are all
determined by the symmetry-breaking scale of $U(1)_{T3R}$, the energy scale of
the $b \bar b \rightarrow \tau \bar \tau$  cross-section is only moderately dependent on
whether the mediating particle is $Z_R$, $\tilde \phi_d$ or $\gamma$.  The
amplitude for $b \bar b \rightarrow \tilde \phi_d \rightarrow \tau \bar \tau$ is thus
proportional to $\lambda_b \lambda_\tau$, while the amplitude for
$b \bar b \rightarrow Z_R \rightarrow \tau \bar \tau$ is proportional to $g_{T3R}^2$.
But the   $b \bar b \rightarrow Z_R \rightarrow \tau \bar \tau$ amplitude can
interfere with the $b \bar b \rightarrow \gamma^* \rightarrow \tau \bar \tau$ amplitude,
enhancing its contribution.
Thus, the rough limits on lepton universality-violating contributions from $\tilde \phi_d$ and $Z_R$ exchange
are
\bea
\lambda_b^2 \lambda_\tau^2 ,\,
g_{T3R}^2 \, g_{em}^2 < 0.001 g_{em}^4,
\eea
where $g_{em}$ is the electromagnetic coupling constant.
For $\tilde \phi$ exchange, we can use eq.~\ref{RHfermmass} to write $\lambda_b$ and
$\lambda_\tau$ in terms of
$\langle \tilde \phi_d \rangle$ and $m_{b,\tau}$.  The flavor constraint can thus be
rewritten as
\bea
\langle \tilde \phi_d \rangle > 50~\gev
\eea
These constraints imply $\lambda_b < 0.1$, $\lambda_\tau < 0.04$, so
the fine-tuning of the bottom and $\tau$ mass terms is reduced by a factor of 5.  More importantly,
it explains the hierarchy which places the $b$ and $c$ quarks and the $\tau$ lepton at
roughly the same mass scale.

\ssection{Dark matter-nucleon scattering cross-section}
In this model, dark matter can scatter off $b$- and $c$-quarks through $t$-channel
exchange of $Z_R$.  Note that since only $b_R$, $c_R$ couples to $Z_R$, the interaction vertex must
have a $V-A$ structure.  In addition, $\eta$ is also chiral under this gauge group, and
couples to $Z_R$ through a $V-A$ interaction vertex.

The most relevant
scattering amplitude is spin-independent, arising from a vector-vector coupling (a
pseudovector-pseudovector spin-dependent coupling may also be present, but will be more
difficult to probe at experiments).  It is easiest to consider the case where the
dark matter particle is a scalar.  In this case, the spin-independent scattering cross-section
is given by
\bea
\sigma_{SI} &=& {m_r^2 \over 4\pi m_{Z_R}^4  } g_{T3R}^4 [ZB_c^p +(A-Z)B_c^n]^2
\eea
where $m_r$=$m_{\tilde \eta } m_N/(m_{\tilde \eta }+m_N)$ and $B_c^{(p,n)} \sim 0.04$~\cite{Ellis:2001hv}.
Since the $Z_R$ mass is generated by symmetry-breaking of $U(1)_{T3R}$, one expects
\bea
\label{mZReq}
m_{Z_R} \sim g_{T3R} \sqrt{\langle \tilde \phi_u \rangle^2 + \langle \tilde \phi_d \rangle^2 }.
\eea
It is worth noting that the interesting region of low-mass
dark matter would correspond to $m_{\tilde \eta} \sim 7-10~\gev$, $g_{T3R} \sim 0.01$ and $m_{Z_R} \sim
1~\gev$ (which is a reasonable choice, given eq.~\ref{mZReq}).\footnote{These
models can potentially match signals from DAMA, CoGeNT and CRESST~\cite{lowmassdata}, but these signals are
seriously challenged by analyses from XENON100, a preliminary analysis from XENON10 and a recent analysis
from CDMS~\cite{lowmassconstraint}}  This is
within the limits imposed by lepton universality. However, for this model, the scattering cross-section can be much lower
since $m_{Z_R}$ can depend on $\langle \tilde \phi_u \rangle$ and other mixing angles.

\ssection{Collider signals}
A standard way to search for dark matter at a hadron collider is by the production
of new colored particles, which then decay to dark matter and Standard Model jets and leptons.
This search strategy is possible in the case of hidden sector asymmetric dark matter,
through QCD production of the exotic quarks,
$pp \rightarrow q_i \bar q_i \rightarrow c \bar c
(b \bar b)\tilde \eta \tilde \eta^*$, where the scalar $\tilde \eta$ is the dark matter
candidate.  This signal is interesting because the production cross-section is controlled
by QCD processes, and thus is independent of $g_{T3R}$ and the Yukawa couplings.
Due to $Z_2$ charge conservation, $q_i$ is constrained to decay to $\tilde \eta$.
The Yukawa coupling $C_i$ only determines the lifetime of $q_i$, and we will assume that $q_i$ decays
within the detector.
As we have seen, the mass of the exotic quarks is not controlled by electroweak
symmetry breaking, so there is no expected maximum scale for $m_{q_i}$.  As such, colliders
cannot exclude this signal.  But if $m_{q_i}$ is within reach of the LHC, then the LHC can
find evidence for this signal, jets plus missing transverse
energy.

This signal may be especially striking in the case where the exotic quark is down-type
and the signature is two $b$-jets and missing $E_T$.
A detailed analysis of this signal is underway,
and preliminary results indicate that the first LHC physics run may be able to probe models
with $m_{q_i} \lsim 600~\gev$~\cite{bjetsmissingET}.
Interestingly, this is also a signal for
WIMPless dark matter.  In that case, the process is pair-production of down-type exotic quarks, which
decay to $b$-quarks and two scalar WIMPless candidates.

Another signal is $pp \rightarrow \tilde b_R \tilde b_R^* \rightarrow b {\bar b} \phi_d \bar \phi_d$,
where $\phi_d$ is the fermionic partner to $\tilde \phi_d$ (i.~e.~, the ``higgsino" of $U(1)_{T3R}$).
Note that $\phi_d$ cannot decay to Standard Model particles.  It
is a fermion which is neutral under $U(1)_{B-L}$, and therefore must decay to an odd number
of fermions for whom $N_B -N_L$ vanish.  Since the MSSM sfermions are much heavier than the
GeV scale, $\phi_d$ cannot decay to any MSSM particles.

$\phi_{u,d}$ is not necessarily a good asymmetric dark matter candidate;
although $m_{\phi_{u,d}} \sim {\cal O}(\gev )$, there is no reason for its number density to be
related to the baryon number density.  Moreover, it may decay to very light hidden sector
particles, with small relic density.  Interestingly, it will still appear as missing
transverse energy at a collider experiment. The lightest particle in the $\eta$ supermultiplet
is still our asymmetric dark matter candidate.

There are other signatures which are similar to Higgs signatures, such as
$pp \rightarrow b\bar b \tilde \phi_d \rightarrow b {\bar b} \tau \bar \tau$ or
$pp \rightarrow \tilde \phi_d \rightarrow \tau \bar \tau$ (with the production
of $\tilde \phi_d$ controlled by a loop of $b$-quarks).  These processes would be somewhat larger
than what is expected for Higgs production, since the $\lambda_{b,\tau}$ Yukawa couplings are larger than the
standard Higgs Yukawas of the $b$ and $\tau$.

\ssection{Conclusions}
We have shown that hidden sector baryogenesis~\cite{Dutta:2006pt} can yield
an asymmetric dark matter candidate which naturally has approximately the
correct relic density.  The dark matter mass $m_\eta$ is set by the mass scale
of the bottom, charm and $\tau$, and thus is $\sim {\cal O}(\gev )$.  This model thus not only
explains why the dark matter and baryon number densities are comparable, but also why
the dark matter relic density is close to  the baryon density.  Interesting tests of this
proposal can be made at the Tevatron and the LHC, where
processes with $b$'s or $\tau$'s in the final
state should be especially amenable to searches at colliders.
The lepton universality-violating process
$\Upsilon \rightarrow \tau \bar \tau$ can potentially be observed at Super-Belle.

\ssection{Acknowledgements}
We are grateful to V.~ Barger, T.~Browder, J.~Feng, F.~Harris, D.~Marfatia,
S.~Pakvasa, A.~Rajaraman, L.~Randall, J.~Rorie and X.~Tata for useful discussions.
This work is supported by  DOE grant
DE-FG02-95ER40917 and DE-FG02-04ER41291.



\end{document}